\documentclass[12pt]{article}
\begin{document}

\begin{center}
{\large \bf About neutral mesons and particle oscillations in the light of field-theoretical prescriptions of Weinberg}

\vspace{0.5 cm}

\begin{small}
\renewcommand{\thefootnote}{*}
L.M. Slad\footnote{slad@theory.sinp.msu.ru} \\
{\it Skobeltsyn Institute of Nuclear Physics,
Lomonosov Moscow State University, Moscow 119991, Russia}
\end{small}
\end{center}

\vspace{0.3 cm}

\begin{footnotesize}
By extending the well-known Weinberg's prescriptions on the diagonalization of the mass term of the Lagrangian without increasing the total number of entities, we get the following conclusions: the set of neutral $K$-mesons consists of two elements, $K_{S}^{0}$ and $K_{L}^{0}$; the states $K^{0}$ and $\bar{K}^{0}$ do not exist as physical objects (in the form of particles or "particle mixtures"); the absence of the states $K^{0}$ and $\bar{K}^{0}$ destroys the grounds for introducing the notion of their oscillations, further replicated as the neutrino oscillation concept. The conclusions concerning the neutral $K$-mesons are also applicable to the neutral $D$-, $B$- and $B_{s}$-mesons. 

\end{footnotesize}

\vspace{0.5 cm}

\begin{small}

\begin{center}
{\large \bf 1. Introduction}
\end{center}

Over a long time, of about 45 years, the monopoly position in solving the solar neutrino problem was occupied by the neutrino oscillation concept. Only recently, an elegant alternative solution to the solar neutrino problem based on logically clear methods of the classical field theory has been presented together with the hypothesis of the existence of semi-weak interaction of electron neutrinos with nucleons caused by the exchange of a massless pseudoscalar boson \cite{1}, \cite{2}. There is a doubtless relevance of an intent analysis of various theoretical aspects of the neutrino oscillation concept. In the present work, we turn to the sources of this concept within the framework of a broader discussion of the existing judgements about neutral $K$-mesons and their revision on the basis of field-theoretical prescriptions of Weinberg \cite {3}.

The necessity in the declared revisions is related to the fact that, as we will be continuously noting, in the situation with neutral $K$-mesons, all elements, being properly ordered, become identical to suitable elements of the Weinberg's electroweak interaction model. A fragments of this model, which we will call the prescriptions or procedure of Weinberg, concerns the diagonalization of the mass term in the Lagrangian, including the transition from the initial classical fields to the final quantum fields with the elimination of the initial fields from the final theory. This procedure seems simple, logical, and natural. It is an exact implementation of a scientific principle existing for hundreds of years that "entities must not be multiplied beyond necessity". Therefore, we accept it as a universal rule of the field theory and particle physics.

Prior to starting a discussion about applying Weinberg's field-theoretical prescriptions to neutral $K$-mesons, we recall the esential motivation and assumptions which had lead to the initial assertions about the family of four neutral $K$-mesons. The assertions that remain nearly unchanged, and one of them, concerning the mutual transitions of $K^{0}$- and $\bar{K}^{0}$-mesons in vacuum, has initially served \cite{4} and continues to serve until now \cite{5}, by analogy, as the only theoretical argument in favor of neutrino oscillation hypothesis.

One of various results of our work is the conclusion on the absence of the mesons $K^{0}$ and $\bar{K}^{0}$ in the nature and, thus, on the absence of any oscillations of neutral mesons. Addressing to the neutrino oscillation concept, we will show its incompatibility with field-theoretical prescriptions of Weinberg and the logical impossibility of its mathematical realization in a number of experimental situations.

\begin{center}
{\large \bf 2. The field-theoretical prescriptions of Weinberg }
\end{center}

For the sake of impeccability of the further conclusions we list the sequence of the separate steps of Weinberg's procedure \cite {3}, despite they are widely known. First step: Weinberg, on the basis of spontaneous breaking of the original symmetry, obtains the mass term in the Lagrangian, which is nondiagonal in the initial gauge fields, $cA_{\mu}^{3}(x)B^{\mu}(x)$ ($c$ is a constant). Second step: on the basis of two suitable superpositions of classical fields $A_{\mu}^{3}(x)$ and $B_{\mu}(x)$,  Weinberg introduces orthonormal classical fields with definite masses $Z_{\mu}(x)$ and $A_{\mu}(x)$. Third step: Weinberg expresses the fields $A_{\mu}^{3}(x)$ and $B_{\mu}(x)$, and then all terms of the gauge Lagrangian, through the fields $Z_{\mu}(x)$ and $A_{\mu}(x)$.  Fourth step: Weinberg gives status of quantized fields to the fields $Z_{\mu}(x)$ and $A_{\mu}(x)$ and identifies them with such particles as the intermediate meson $Z$ and the photon. First note: at any stage of constructing the gauge model, Weinberg does not connect the fields $A_{\mu}^{3}(x)$ and $B_{\mu}(x)$ with any quanta. Second note: the original gauge field $A_{\mu}^{3}(x)$ and $B_{\mu}(x)$ that serve as the cornerstone in the Weinberg's construction disappear in the final model of electroweak interactions. Third note: the initial fields possess well-defined values of weak isospin and its third projection, but the final fields $Z_{\mu}(x)$ and $A_{\mu}(x)$ do not have such definite values, and, therefore, the corresponding terms of the electroweak interaction Lagrangian violate the isospin symmetry.

\begin{center}
{\large \bf 3. The origin of the existing judgments about neutral $K$-mesons}
\end{center}

We now turn to a number of current opinions about the neutral $K$-mesons. They mainly reproduce part of the judgments stated in the works by Gell-Mann \cite{6} and Gell-Mann and Pais \cite{7} with adding some corrections for the results of the subsequent experiments concerning the violation of $CP$-symmetry.

Starting from the assumption of strict conservation of the isotopic spin in strong interactions, Gell-Mann \cite{6} concludes that, among the two long-lived neutral particles produced in the collision of the $\pi^{-}$meson with the proton, one particle ($K^{0}$) is a boson with the isospin 1/2 and its projection -1/2 and that there exists an antiparticle ($\bar{K}^{0}$) with the isospin projection +1/2 which corresponds to the boson $K^{0}$ and is  different from it. The mentioned assumption of Gell-Mann is the key element in the subsequently formed structure of the family of neutral $K$-mesons. In the picture of neutral $K$-meson, which we represente here, such an assumption is incorrect.

Gell-Mann and Pais \cite{7} consider that, if there exists the decay $K^{0} \rightarrow \pi ^{+}+\pi^{-}$, then there should exist the charge-conjugate process $\bar{K}^{0} \rightarrow \pi ^{+}+\pi^{-}$, and thus, the weak interaction causes the virtual transition $K^{0} \rightleftharpoons \pi^{+}+\pi^{-} \rightleftharpoons \bar{K}^{0}$. The last judgment and the aspiration for providing the $C$-parity conservation in weak decays lead to the introduction of the quanta $K_{1}^{0}$ and $K_{2}^{0}$, which fields are expressed in the form of normalized sum and difference of the fields $K^{0}$ and $\bar{K}^{0}$, respectively. According to Gell-Mann and Pais, $K^{0}$ and $\bar{K}^{0}$ are the primary objects in production phenomena, whereas the decay process is best described in terms of $K_{1}^{0}$ and $K_{2}^{0}$. Each of the latter can be assigned a definite lifetime, that is not true to the $K^{0}$ and $\bar{K}^{0}$.

Note that a long-lived neutral boson produced, for example, in the collision of a $\pi^{-}$-meson with a proton, cannot make any experimental manifestation between the procuction moment and the decay moment and, consequently, it does not allow experimental identification in this time interval. The opinion stated in \cite{6} and \cite{7} that such a boson must have a well-defined value of the isospin and its third projection remains nothing more but an assumption. The fact that the introduced hadrons $K_{1}^{0}$ and $K_{2}^{0}$ do not possess definite values of the third projection of the isospin, does not induce Gell-Mann and Pais to reconsider their assumption of strict isotopic spin conservation in strong interactions, and this essentially prohibits the participation of these hadrons in strong interactions, causing at least a surprise.

Essential is the position of the authors of work \cite{7} about reserving the word "particle" for an object with a definite lifetime and recognizing the quanta $K_{1}^{0}$ and $K_{2}^{0}$ as true "particles", and about that the $K^{0}$ and $\bar{K}^{0}$ must, strictly speaking, be considered as "particle mixtures" is presented essential. The mathematical definition of "particle mixtures" has resulted by Pais and Piccioni \cite{8} in introducing and describing the concept of oscillations, mutual transitions of the $K^{0}$ and $\bar{K}^{0}$-mesons in vacuum.

Let us make a brief summary. At Weinberg's construction of the electroweak interaction model, two neutral classical (unquantized) fields were the original elements, while at the construction of the $K$-meson picture, two neutral quantized objects were the original elements. The opinion on the existence of a virtual transition $K^{0} \rightleftharpoons \pi^{+}+\pi^{-} \rightleftharpoons \bar{K}^{0}$ reflects in fact the existence of a mass term in the Lagrangian, which is nondiagonal in the respective fields. The situation with doubling the number of neutral $K$-mesons by saving the two ptimary and adding two new mesons can, in our opinion, be classified as the creation of a logical chaos. This chaos does not allow a simply connected mathematical description of the evolution of any neutral $K$-meson including its production and decay. To link the disparate parts of the mathematical description together, one uses  judgments which are rather verbal than based on the Lagrangian or any equation. The verbal judgment is, for example, that the $K^{0}$-meson turns after its production into an object described by a superposition of particles with different masses, $K_{1}^{0}$ and $K_{2}^{0}$. The elimination of the infinite arbitrariness in the choice of states of these particles requires a new verbal judgment.

\begin{center}
{\large \bf 4. Utilization of Weinberg's field-theoretical prescriptions for neutral $K$-mesons}
\end{center}

Following Weinberg, we shall deal initially not with particles but with suitable classical fields and assume that the initial Lagrangian of strong interactions is invariant under transformations of the isospin group $SU(2)$ (or the internal symmetry group $SU(3)$).

As the first step, we fix the fields and establish the presence of a nondiagonal mass term in the Lagrangian of these fields. Let's introduce two neutral fields $\Phi_{+1 \frac{1}{2} -\frac{1}{2}}(x)$ and $\Phi_{-1 \frac{1}{2} +\frac{1}{2}}(x)$ that are pseudoscalar under the orthochronous Lorentz group and possess strangenesses $\pm 1$, isospin $1/2$ and its projections $\mp 1/2$ (these fields can also be considered as the components of vectors in the octet representation space of the group $SU(3)$). We assume at the stage of preliminary analysis that, in the absence of weak interactions, these fields could describe the lower bound states of quark-antiquark systems, respectively, $d\bar{s}$ and $s\bar{d}$. In the presence of weak interactions, these two systems would virtually pass into one another due to double exchange of $W$-bosons with changing the third projection of isospin and strangeness. (Feynman diagrams corresponding to such an exchange can be found in the monograph \cite{9}). This means that the mass term in the Lagrangian of fields $\Phi_{\pm 1 \frac{1}{2} \mp \frac{1}{2}}(x)$, in view of the influence of weak interactions, should be presented in the following form
$${\cal L} = -m_{+}^{2}\Phi_{+1\frac{1}{2} -\frac{1}{2}}^{*}(x)
\Phi_{+1\frac{1}{2} -\frac{1}{2}}(x)
-m_{-}^{2}\Phi_{-1\frac{1}{2} +\frac{1}{2}}^{*}(x)
\Phi_{-1\frac{1}{2} +\frac{1}{2}}(x)$$
\begin{equation}
-a\Phi_{+1\frac{1}{2} -\frac{1}{2}}^{*}(x)\Phi_{-1\frac{1}{2} +\frac{1}{2}}(x)
-a^{*}\Phi_{-1\frac{1}{2} +\frac{1}{2}}^{*}(x)
\Phi_{+1\frac{1}{2} -\frac{1}{2}}(x),
\label{1}
\end{equation}
where $m_{+}$ and $m_{-}$ are real constants.

If the mass term of Lagrangian (\ref{1}) does not change under the transformation $\Phi_{\pm 1 \frac{1}{2} \mp \frac{1}{2}}(x)\rightarrow \Phi_{\mp 1 \frac{1}{2} \pm \frac{1}{2}}(x)$ (it is possible to consider this term as possessing $SU(3)$-symmetry) then the constant $a$ is real, and the quantities $m_{+}^{2}$ and $m_{-}^{2}$ are equal. Its diagonalization, as the second step of the appropriate procedure of Weinberg, leads to the fields $\Phi_{1}(x)$ and $\Phi_{2}(x)$ with certain masses given by the expressions
\begin{equation}
\Phi_{1}(x) = \frac{1}{\sqrt{2}}[\Phi_{+1\frac{1}{2} -\frac{1}{2}}(x)+
\Phi_{-1\frac{1}{2} +\frac{1}{2}}(x)], \quad
\Phi_{2}(x) = \frac{1}{\sqrt{2}}[\Phi_{+1\frac{1}{2} -\frac{1}{2}}(x)-
\Phi_{-1\frac{1}{2} +\frac{1}{2}}(x)],
\label{2}
\end{equation}
whose form is the same for all nonzero values of the constants $a$ and $m_{\pm}^{2}$. The definite values of the $CP$-parity of fields $\Phi_{1}(x)$ and $\Phi_{2}(x)$ are the result and not the ground for the formation of expressions (\ref{2}).

Since the experiments indicate absence of definite values of $CP$-parity of the decaying neutral $K$-mesons, one should consider the case when the mass term in Lagrangian (\ref{1}) does not possess the mentioned $SU(3)$-invariance. Then its diagonalization gives orthonormal fields $\Phi_{S}(x)$ and $\Phi_{L}(x)$ with definite masses $m_{S}$ and $m_{L}$ specified by the formulas
\begin{equation}
\Phi_{S}(x) = \frac{(1-\varepsilon^{*})\Phi_{+1\frac{1}{2} -\frac{1}{2}}(x)+
(1+\varepsilon^{*})\Phi_{-1\frac{1}{2} +\frac{1}{2}}(x)}
{\sqrt{2(1+|\varepsilon|^{2})}} =
\frac{\Phi_{1}(x)-\varepsilon^{*}\Phi_{2}(x)}{\sqrt{1+|\varepsilon|^{2}}},
\label{3}
\end{equation}
\begin{equation}
\Phi_{L}(x) = \frac{(1+\varepsilon)\Phi_{+1\frac{1}{2} -\frac{1}{2}}(x)-
(1-\varepsilon)\Phi_{-1\frac{1}{2} +\frac{1}{2}}(x)}
{\sqrt{2(1+|\varepsilon|^{2})}} =
\frac{\varepsilon \Phi_{1}(x)+\Phi_{2}(x)}{\sqrt{1+|\varepsilon|^{2}}},
\label{4}
\end{equation}
with
\begin{equation}
|m_{L}^{2}-m_{S}^{2}| = \sqrt{(m_{+}^{2}-m_{-}^{2})^{2}+4|a|^{2}},
\label{5}
\end{equation}
\begin{equation}
\varepsilon = \frac{m_{+}^{2}-m_{-}^{2}+2i{\rm Im} a}
{m_{L}^{2}-m_{S}^{2}-2{\rm Re} a}.
\label{6}
\end{equation}

The next step of the Weinberg's procedure consists in finding expressions for the fields $\Phi_{\pm 1 \frac{1}{2} \mp \frac{1}{2}}(x)$ through the fields $\Phi_{S}(x)$ and $\Phi_{L}(x)$ on the basis of the relations(\ref{3}) and (\ref{4}) and in substituting these expressions into all terms in the initial Lagrangian describing both strong and weak interactions.

Then, in completing the Weinberg's procedure, we should obtain Euler equations for each of the fields $\Phi_{S}(x)$ and $\Phi_{L}(x)$ from the transformed Lagrangian and perform the second quantization of the solutions of these equations, that would consist in identifying these solutions as vectors in the spaces of suitable irreducible representations of the Poincare group with neutral mesons $K_{S}^{0}$ and $K_{L}^{0}$. As a result, the processes which were thought to be accompanied by the production or are caused by an interaction of one of the hypothetical $K^{0}$- and $\bar{K}^{0}$-bosons, with never being accompanied with the production or caused by an interaction of other one, can now proceed within the considered concept with involving both
$K_{S}^{0}$- and $K_{L}^{0}$-bosons with almost equal probabilities (though, slightly different probabilities, that is important for a number of experiments). So, the production of $\Lambda$-hyperon in a $\pi^{-}p$-collision
is accompanied with either $K_{S}^{0}$- or $K_{L}^{0}$-meson, with both of the latter being able to produce $K^{-}$-meson (when interacting with neutrons) and $K^{+}$-meson (when interacting with protons).
  
Fields $\Phi_{\pm 1 \frac{1}{2} \mp \frac{1}{2}}(x)$ and $\Phi_{1,2}(x)$, which are involved in mathematical operations to form the fields $\Phi_{S}(x)$ and $\Phi_{L}(x)$ and, that is especially important, to form the $K_{S}^{0}$- and $K_{L}^{0}$-mesons interaction constants, are not identified with any quanta at any step of such a formation.

\begin{center}
{\large \bf 5. The picture of neutral mesons in the light of Weinberg's prescriptions}
\end{center}

Thus, the postulated universality of the Weinberg's prescriptions on the diagonalization of the mass term in the Lagrangian without increasing the total number of entities leads to the following conclusions:

(1) The states $K^{0}$ and $\bar{K}^{0}$, as physical objects (in the form of particles or "particle mixtures"), do not exist. Accordingly, the quark-antiquark bound states $d\bar{s}$ and $s\bar{d}$ are not realized;

(2) The set of neutral $K$-mesons consists of two elements, $K_{S}^{0}$ and $K_{L}^{0}$, which do not possess definite values of the third projection of the isospin. Accordingly, the bound states are formed only by the superpositions of quark-antiquark pairs $d\bar{s}$ and $s\bar{d}$ which are given by the expressions, obtainable in an obvious way from the formulas (\ref{3}) and (\ref{4});

(3) In strong interactions involving neutral $K$-mesons, $K_{S}^{0}$ and $K_{L}^{0}$, the isospin is not conserved;

(4) The absence of the states $K^{0}$ and $\bar{K}^{0}$ destroys the ground for introducing the notion of their oscillations made by Pais and Piccioni \cite{6}.

All the listed conclusions about the family of neutral $K$-mesons should, in our view, also be extended to the sets of neutral $D$-, $B$- and $B_{s}$-mesons. In particular, each of these sets consists of two elements, namely, 
$D^{0}_{1}$ and $D^{0}_{2}$, $B^{0}_{H}$ and $B^{0}_{L}$, $B^{0}_{sH}$ and $B^{0}_{sL}$, and the states $D^{0}$ and $\bar{D}^{0}$, $B^{0}$ and $\bar{B}^{0}$, $B^{0}_{s}$ and $\bar{B}^{0}_{s}$ do not exist.

\begin{center}
{\large \bf 6. About some features of the neutrino oscillation concept}
\end{center}

The outlined concept of neutral mesons, precisely following the Weinberg's prescriptions on the transition from the initial gauge fields to the observed particles, is simple and consistent both in the theoretical and experimental aspects. None of its parts can serve as an analogy or basis for the hypothesis of neutrino oscillations. The radical difference between the judgements on the family of neutral $K$-mesons and the present day dominanting judgements on the family of neutrinos start with the origin of nondiagonal mass terms in the Lagrangian of the initial fields and are strengthened by the total number of entities. Namely, such terms in the Lagrangian of neutral meson fields with isospin 1/2 are necessitated, as it was indicated above, by weak interactions, and the number of the initial and final meson fields remains the same. At the same time, the nondiagonal mass terms in the Lagrangian of the known neutrino fields are only arbitrarily introduced (see, e.g., \cite{10}) to justify adding new entities, the massive neutrinos $\nu_{j}$, $j=1,2,3$, as "true particles", to justify declaring the known neutrinos $\nu_{\alpha}$, $\alpha = e, \mu, \tau$, as "particle mixtures"  and to realize the oscillation scenario a la Pais and Piccioni. It is worth noting that the "true particles" $\nu_{j}$ do not directly participate in any production, scattering or absorption processes.

The redundancy of neutrino family inevitably leads to logical contradictions and uncertainties. Attempts to avoid some of the contradictions in the description of neutrino oscillations can be found, e.g., in the works \cite{11} and \cite{12}.

We would now like to draw attention to only one essential logical aspect related to the concept of neutrino oscillations. Note in the beginning, that the realization of the notion of "particle mixtures", since the times of Pais and Piccioni \cite{8} up to present day (see, e.g., \cite{4}), contains the elements both quantum and classical mechanics. With respect to massive neutrino states whose superpositions correspond to the states describing the production of any of the known neutrinos $\nu_{\alpha}$, $\alpha = e, \mu, \tau$, ones adopt the assumption (see, e.g.,\cite{4}), that each of them is described by its well-defined value of 4-momentum. Let us briefly elucidate the possible consequences of non-fulfilment of the specified assumptions.

It is accepted that the state vector $|\nu_{\alpha}(t_{0},{\bf r})\rangle$ of the neutrino $\nu_{\alpha}$, if exists, does not change in time, and that at the time $t_{0}$ of the neutrino production it is linearly expressed through the state vectors of all neutrinos $\nu_{j}$
\begin{equation}
|\nu_{\alpha}(t_{0},{\bf r})\rangle = \sum_{j=1,2,3} u_{\alpha j}
|\nu_{j}(t_{0},{\bf r})\rangle, \qquad \alpha = e, \mu, \tau,
\label{7}
\end{equation}
where the coefficients $u_{\alpha j}$ do not depend on the production conditions of the neutrino $\nu_{\alpha}$. As the state vector $|\nu_{j}(t_{0},{\bf r})\rangle$ belongs to the space of irreducible representations of the Poincare group characterized by the mass $m_{j}$, its dependence on the spatial coordinates determines the distribution of the state over 3-momentum and thus specifies the evolution in time
\begin{equation}
|\nu_{j}(t,{\bf r})\rangle = \int |\nu_{j}({\bf p})\rangle 
\exp[-i\sqrt{{\bf p}^{2}+m_{j}^{2}}(t-t_{0})+i{\bf p}{\bf r}] d^{3}{\bf p}.
\label{8}
\end{equation}

Let the state of an evolving in time quantum-mechanical object identified at its production moment $t_{0}$ with neutrinos $\nu_{\alpha}$ be denoted as $|Z_{\alpha}(t,{\bf r})\rangle$: $|Z_{\alpha}(t_{0},{\bf r})\rangle = |\nu_{\alpha}(t_{0},{\bf r})\rangle$. On one hand, for any time $t > t_{0}$ it is natural to describe this state by a time-dependent superposition of states presented in the right part of relation (\ref{7})
\begin{equation}
|Z_{\alpha}(t,{\bf r})\rangle = \sum_{j=1,2,3} u_{\alpha j}
|\nu_{j}(t,{\bf r})\rangle.
\label{9}
\end{equation}
On the other hand, the essence of the concept of neutrino oscillations is that the state $|Z_{\alpha}(t,{\bf r})\rangle$, $\alpha = e, \mu, \tau$, is expressed as superposition of the three states $|\nu_{\beta}(t_{0},{\bf r})\rangle$
\begin{equation}
|Z_{\alpha}(t,{\bf r})\rangle = \sum_{\beta = e, \mu, \tau} v_{\alpha \beta}(t)
|\nu_{\beta}(t_{0},{\bf r})\rangle,
\label{10}
\end{equation}
where $v_{\alpha \beta}(t)$ are some $c$-number valued functions of time.

From relations (\ref{9}), (\ref{8}) and (\ref{7}), we conclude that the equality (\ref{10}) is feasible if and only if each of the three states $|\nu_{j}(t,{\bf r})\rangle$ (\ref{8}) is characterized by its definite the 3-momentum absolute value $|{\rm p}_{j}|$, $j=1,2,3$. If such a condition concerning the distributions of the massive neutrino states over 3-momentum is not fulfilled, that is likely enough from the field-theoretical point of view, then the realization of the concept of neutrino oscillations in such a situation becomes impossible, namely, the states of $|Z_{\alpha}(t,{\bf r})\rangle$ for $t > t_{0}$ are not representable as superpositions of the neutrino states $\nu_{e}$, $\nu_{\mu}$ and $\nu_{\tau}$. Because of that, we lose the possibility to calculate the rate of any expected process involving the object $|Z_{\alpha}(t,{\bf r})\rangle$.

\begin{center}
{\large \bf 7. Conclusion}
\end{center}

The recognition of the identity between a number of initial theses in  Weinberg's construction of the electroweak interaction model and in introducting the notion of neutral bosons with isospin 1/2 entails the necessity in extending Weinberg's prescriptions to the construction of a picture of neutral $K$-mesons without extra entities. The presented picture of neutral mesons radically differs from the one dominating now by its physical and logical simplicity.

I am sincerely grateful to S.P. Baranov, M.Z. Iofa, A.E. Lobanov, A.M. Snigirev and S.P. Volobuev for the numerous discussions of problems connected with the present work.

\end{small}

\begin{thebibliography}{99}

\bibitem{1}
   L.M. Slad, preprint, arXiv:1502.03262v3 [hep-ph], 2018. 
\bibitem{2}
   L.M. Slad, preprint, arXiv:1808.05103v1 [hep-ph], 2018.
\bibitem{3}
   S. Weinberg, Phys. Rev. Lett. 19 (21) (1967) 1264-1266.
\bibitem{4}
   B. Pontecorvo, Sov. Phys. JETP 7 (1) (1958) 172-173.
\bibitem{5}
   K. Nakamura, S.T. Petcov, in: C. Patrignani et al. (Particle Data Group), Chin. Phys. C 40 (2016) 100001.   
\bibitem{6}
   M. Gell-Mann, Phys. Rev. 92 (3) (1953) 833-834.
\bibitem{7}
   M. Gell-Mann, A. Pais, Phys. Rev. 97 (5) (1955) 1387-1389.
\bibitem{8}
   A. Pais, O. Piccioni, Phys. Rev. 100 (5) (1955) 1487-1489.
\bibitem{9}
   L.B. Okun, Leptons and Quarks, North-Holland, Amsterdam, 1982.
\bibitem{10}
   V. Gribov, B. Pontecorvo, Phys. Lett. B 28 (7) (1969) 493-496.
\bibitem{11}
   B. Kayser, Phys. Rev. D 24 (1) (1981) 110-116.
\bibitem{12}
   C. Giunti, C.W. Kim, U.W. Lee, Phys. Rev. D 44 (11) (1991) 3635-3640.

\end{thebibliography}
\end{document}